\documentclass[aps,prb,twocolumn,superscriptaddress,floatfix]{revtex4}
\usepackage{graphicx}
\usepackage{epsfig}
\usepackage{amsmath}
\usepackage{amssymb}
\usepackage{subfigure}
\usepackage{relsize}
\usepackage{mathptmx}

\begin{document}

\frenchspacing

\title{The Ag$\ldots$Ag dispersive interaction and exotic physical properties of Ag$_3$Co(CN)$_6$}

\author{Hong Fang}
\affiliation{Department of Earth Sciences, University of Cambridge, Downing Street, Cambridge, CB2 3EQ, U.K.}

\author{Martin T. Dove}
\email{martin.dove@qmul.ac.uk}
\affiliation{Department of Earth Sciences, University of Cambridge, Downing Street, Cambridge, CB2 3EQ, U.K.}
\affiliation{Centre for Condensed Matter and Materials Physics, School of Physics and Astronomy, Queen Mary University of London, Mile End Road, London, E1 4NS, U.K.}
\affiliation{Materials Research Institute, Queen Mary University of London, Mile End Road, London, E1 4NS, U.K.}

\author{Keith Refson}
\affiliation{Science and Technology Facilities Council, Rutherford Appleton Laboratory, Harwell Science and Innovation Campus, Didcot, Oxfordshire, OX11 0QX, UK}

\date{\today}

\begin{abstract}
We report a density functional theory (DFT) study of Ag$_3$Co(CN)$_6$, a material noted for its colossal positive and negative thermal expansion, and its giant negative linear compressibility. Here we explicitly include the dispersive interaction within the DFT calculation, and find that it is essential to reproduce the ground state, the high-pressure phase, and the phonons of this material; and hence essential to understand this material's remarkable physical properties. New exotic properties are predicted. These include heat enhancement of the negative linear compressibility, a large reduction in the coefficient of thermal expansion on compression with change of sign of the mode Gr\"{u}neisen parameters under pressure, and large softening of the material on heating. Our results suggest that these are associated with the weak Ag$\ldots$Ag dispersive interactions acting with an efficient hinging mechanism in the framework structure.
\end{abstract}


\maketitle

\section{Introduction}

\begin{figure}[t]
\begin{center}
\includegraphics[width=0.5\textwidth]{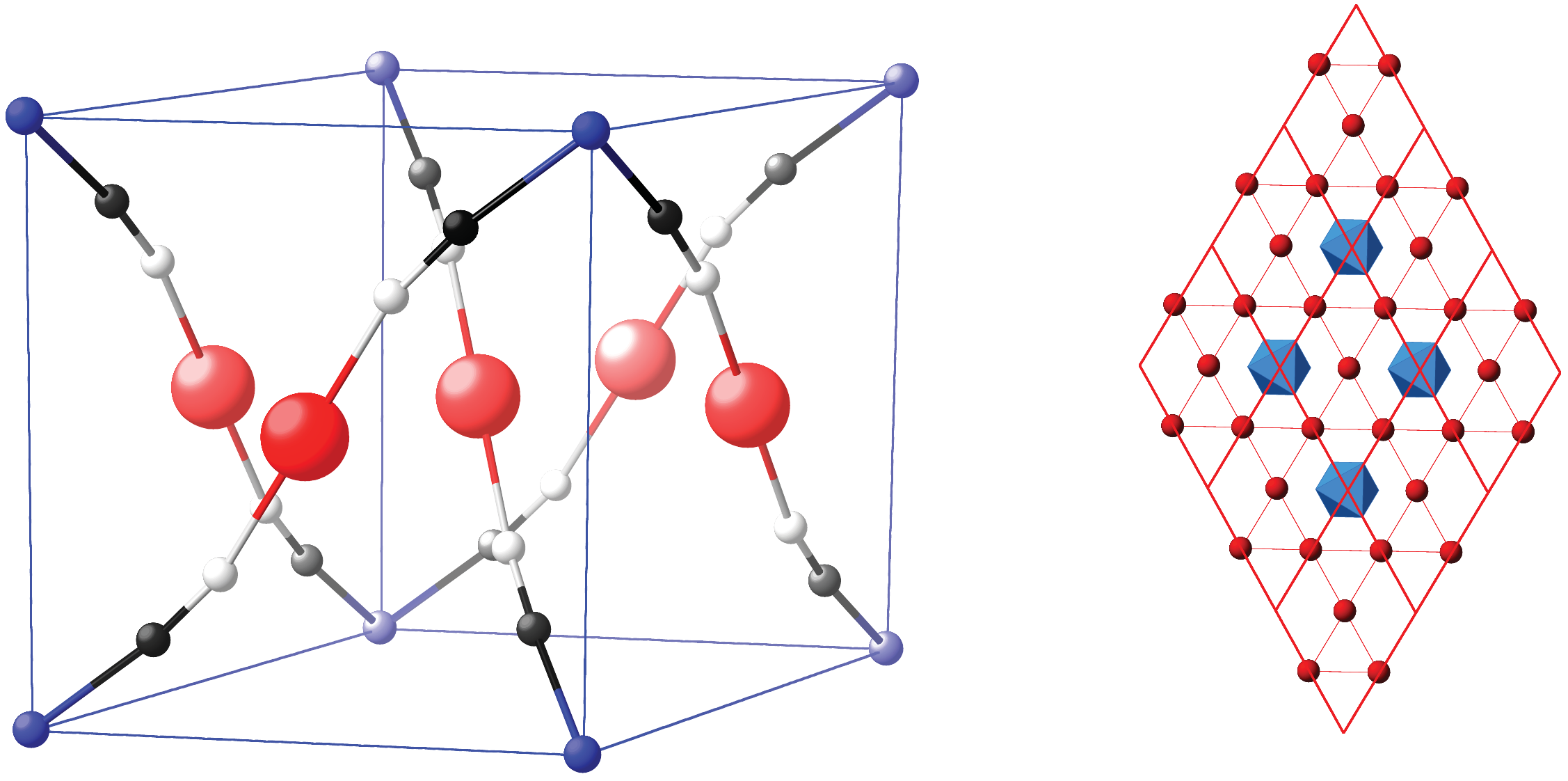}
\end{center}
\caption{\label{fig:primitive} Ambient phase $P\bar{3}1m$ of Ag$_3$Co(CN)$_6$: (a) unit cell with silver in red, cobalt in blue, carbon in black and nitrogen in white grey; (b) looking down the $\left[0,\,0,\,1\right]$ direction with Ag atoms (red) in a Kagome sheet connected to the octahedra [Co(CN)$_6$]$^{3-}$ anions (blue) above and below.}
\end{figure}

Ag$_3$Co(CN)$_6$ has attracted a lot of attention due to its colossal positive and negative thermal expansion~\cite{GoodwinAgCoCN2008,Conterio2008}, and also because of its giant negative linear compressibility~\cite{GoodwinAgCoCNnlc2008}. The negative thermal expansion (NTE) along the $c$-axis and the positive thermal expansion (PTE) along the $a$($b$) axes are an order of magnitude larger than that observed in many other crystalline solids. The material also shows negative linear compressibility (NLC), namely  along the $c$-axis, that is several times greater than the typical value found in crystals. As shown in Fig.~\ref{fig:primitive}, the ambient-pressure phase of Ag$_3$Co(CN)$_6$ has a trigonal structure with space group $P\bar{3}1m$. The structure consists of layers of Kagome sheets of Ag atoms in the $(001)$ crystal plane at height $z=1/2$, with Co--CN--Ag--NC--Co chains along the $\langle011\rangle$ lattice directions linking  [Co(CN)$_6$]$^{3-}$ octahedra. These chains are hinged together in a way that gives the structure a high degree of flexibility; expansion in the trigonal $(001)$ plane is accompanied by a shrinkage in the orthogonal direction in a way that does not change the relevant bond lengths.

Previous \emph{ab initio} density functional theory (DFT) calculations were unable to reproduce the correct ground-state structure and the high-pressure phase of the material~\cite{Calleja2008,Hermet2013,Mittal2012}. Whilst these studies were able to reproduce the lengths of the Co--C, C--N and N--Ag bonds which characterise the structure, the predicted lattice parameters differ considerably from the experimental values. The key interatomic distance that changes as the structure flexes is the Ag\ldots Ag distance, which is equal to half the value of the $a$ lattice parameter. The first of the DFT studies~\cite{Calleja2008} showed that a \textit{post hoc} correction for dispersive interactions between the Ag cations was sufficient to shift the equilibrium DFT structure into good agreement with the experimental crystal structure. The same study also showed that there is no significant covalent bonding between neighbour Ag atoms; it was this factor, combined with the fact that DFT calculations on the structural analogue in which hydrogen or deuterium atoms replace the Ag atoms are in excellent agreement with experiment, that suggested an important role for dispersive Ag$\ldots$Ag interactions.

On this basis, it would be useful to see if a DFT calculation that explicitly includes a correction for the long-range dispersive forces will reproduce the ground state and the high-pressure phase of Ag$_3$Co(CN)$_6$ correctly. If so, it should then be possible to obtain reliable phonons via such calculation in order to better understand the exotic behaviour of this material.

Modern implementations of DFT now include a correction for the long-range dispersive interactions~\cite{Grimme2004,Grimme2006,Dobson2006,Dion2004,Thonhauser2007,Tkatchenko2009,Roman-Perez2009}. One widely used method is called `DFT+D2'~\cite{Grimme2006} where a dispersive interaction that is dampened at short range to avoid double counting of energy is added to the DFT energy from the generalised-gradient approximation (GGA) calculation. Semi-empirical parameters in such a dispersive interaction are provided in Ref.~\onlinecite{Grimme2006} for most elements in the periodic table. The method has been successfully applied to various materials in which the dispersive interactions are important. One good example is the recent work on cesium halides by Zhang et al.~\cite{Zhang2013}, where the DFT+D2 formalism gives both an improved agreement between the optimised and experimental crystal structures and a correct prediction of the ground-state phases.

In this work, we have carried out DFT+D2 calculations for Ag$_3$Co(CN)$_6$. This has confirmed that the inclusion of dispersive forces give the correct ground state structure, as anticipated in the first DFT study of this material~\cite{Calleja2008}. It is also shown that the DFT+D2 model correctly gives the structure of the high-pressure phase; without the dispersive interaction DFT gives a structure without the interdigitation found experimentally~\cite{GoodwinAgCoCNnlc2008}. On the basis of these successes it is now reasonable to investigate the lattice dynamics of Ag$_3$Co(CN)$_6$, from which we have been able to study a number of physical and thermodynamic properties. These form the focus of this paper.

\section{Methods}

\subsection{DFT calculations}

The DFT calculations were performed using the CASTEP code~\cite{Segafll2002}. For comparison, we used both local-density approximation (LDA) and GGA of Perdew-Burke-Ernzerhof (PBE)~\cite{Perdew1996} for the exchange-correlation functional. Optimized norm conserving pseudopotentials generated using the RRKJ method~\cite{Rapper1990} as implemented in the OPIUM package and with parameters from the Rappe and Bennett library~\cite{link} were used in various calculations. A plane-wave basis set was used with the cut-off energy of 1800~eV. Sampling of the Brillouin zone was performed on a $6\times6\times6$ Monkhorst-Pack (MP)~\cite{Monkhorst1976} grid.

The geometries of all structures were optimised using the BFGS method to achieve a convergence of less than $10^{-6}$ eV per atom change in energy per cycle and a force residual of $5\times10^{-4}$~eV/{\AA}. At different pressures, tolerance for accepting convergence of the maximum stress component during unit cell optimization is $5\times10^{-3}$ GPa.

\begin{table*}[t]
\caption{\label{tab:groundstate} Calculated ground-state structures (from GGA+D, GGA and LDA), including the unit-cell edges ($a=b$ and $c$), fractional coordinates of C and N, and the nearest-neighbouring ion distances. $V$ is the volume of one formula unit (note that there is one formulate unit per unit cell). The Ag--Ag distance is equal to $a/2$. $\Delta_\mathrm{GGA+D}$, $\Delta_\mathrm{GGA}$ and $\Delta_\mathrm{LDA}$ represent the deviations of the different calculations compared to experiment at a temperature of 10~K from reference \onlinecite{Conterio2008}.}
\begin{tabular}{@{\extracolsep{8pt}}c|ccc|c|ccc}
\hline  & LDA & GGA & GGA+D & Experiment & $\Delta_\mathrm{GGA+D}$  & $\Delta_\mathrm{GGA}$& $\Delta_\mathrm{LDA}$ \\
\hline
$a\left(=b\right)$ (\AA) & $6.118$ & $7.629$ & $6.664$ & $6.754$ & $-1.3$\% & $+13$\% & $-9$\% \\
$c$ (\AA)                & $7.626$ & $6.621$ & $7.416$ & $7.381$ & $+0.5$\% & $-10$\% & $+3$\%\\
$V$ (\AA$^3$)           & $247.2$ & $333.7$ & $285.2$ & $291.6$  & $-2$\% & $+14$\% & $-15$\% \\
C$_x$                      & $0.238$ & $0.202$ & $0.225$ & $0.220$ & $+0.003$ & $-0.020$  & $+0.016$ \\
C$_z$                      & $0.154$ & $0.171$ & $0.158$ & $0.153$ & $+0.002$ & $+0.015$  & $-0.002$ \\
N$_x$                      & $0.364$ & $0.321$ & $0.347$ & $0.342$ & $+0.008$ & $-0.018$  & $+0.025$ \\
N$_z$                      & $0.269$ & $0.282$ & $0.270$ & $0.266$ & $+0.006$ & $+0.018$  & $+0.005$ \\
C--N (\AA)               & $1.170$ & $1.164$ & $1.164$ & $1.170$ & $-0.5$\% & $-0.5$\%  & $0$\% \\
Ag--N (\AA)              & $1.948$ & $1.988$ & $1.983$ & $2.034$ & $-2.5$\% & $-2.3$\%  & $-4$\%\\
Co--C (\AA)              & $1.868$ & $1.914$ & $1.906$ & $1.865$ & $+2.2$\% & $+2.6$\%  & $+0.2$\%\\
\hline
\end{tabular}
\end{table*}

\subsection{DFT+D2 calculations}

The dispersive contribution was directly added to the DFT GGA energy using a semi-empirical form introduced by Grimme~\cite{Grimme2006},
\begin{eqnarray}\label{g06}
E_\mathrm{disp}  =  - s_6 \sum\limits_{i = 1}^{N - 1} {\sum\limits_{j = i + 1}^N {\frac{{C_6^{ij} }}{{R_{ij}^6 }}} } f_\mathrm{damp} \left( {R_{ij} } \right)
\end{eqnarray}

\noindent where $N$ the number of atoms in the system. $C_6^{ij}$ is the dispersion coefficient of atomic pair $\left(i,j\right)$ that can be computed from the dispersion coefficient of the individual atoms as
\begin{eqnarray}\label{dispcoefficient}
C_6^{ij}  = \sqrt {C_6^i C_6^j }
\end{eqnarray}

\noindent where $R_{ij}$ is the distance between the two atoms, and $R_r$ is the sum of the atomic van der Waals radii of the pair. The dampening factor $f_\mathrm{damp}$ is defined as
\begin{eqnarray}\label{dampenfactor}
f_\mathrm{damp} \left( {R_{ij} } \right) = \frac{1}{{1 + \exp \left[ { - d\left( {R_{ij} /R_r  - 1} \right)} \right]}}
\end{eqnarray}

\noindent with $d=20$. $s_6$ is a scaling factor dependent on the functional used in the calculation; for PBE, $s_6=0.75$. This method has been implemented in CASTEP for geometry optimisation. In what follows we will refer to this method as `GGA+D'; calculations without the dispersion correction will simply be labelled as `LDA' or `GGA' as appropriate.

\subsection{Lattice dynamics with DFPT+D2}\label{section:DFTPD2methods}

Density functional perturbation theory (DFPT)~\cite{Baroni2001,Refson2006} was used to calculate phonons on a $5\times5\times5$ grid of wave vectors, and frequencies for phonons of other wave vectors were then obtained using interpolation~\cite{Baroni2001}. Phonon density of states (DoS) were calculated using a $25\times25\times25$ MP grid~\cite{Monkhorst1976} corresponding to a total of 1470 independent wave vectors.

At the present time CASTEP can only support a DFT+D2 calculation for phonons using the supercell method of finite displacement~\cite{Karki1997}, which turns out to be too expensive to be feasible for Ag$_3$Co(CN)$_6$. Therefore, we first carried out a regular DFPT phonon calculation using CASTEP to get the corresponding dynamical matrices of different wave vectors. We then used the dispersive interaction of Eq.~\eqref{g06} implemented in the lattice simulation program GULP~\cite{Gale1997} to calculate its contribution to the dynamical matrices separately, all based on the same optimised structure from GGA+D. The dynamical matrices from the two codes are added together using a combination of Python scripts and the use of MATLAB, and the combined dynamical matrix was diagonalised to give the  phonon frequencies with effects of the dispersive interaction included. For future convenience, we call this the `DFPT+D' method.

To check the accuracy of our scripts for the DFPT+D method, we performed a benchmark phonon calculation for NaI, chosen because it has a large refractive index (the largest among alkali halides~\cite{Li1976}) and hence likely to have a significant dispersive energy term. This material has a simple structure with only 2 atoms in the primitive cell, so that it was feasible to carry out a DFT+D2 phonon calculation using the supercell method in CASTEP (here called the `supercell+D' method). By comparing the calculated phonon frequencies from DFPT+D and supercell+D, we found the two agree with each other extremely well, with a mean relative discrepancy less than $2\%$ (see phonon dispersion curves in the Supplemental Material~\cite{supplemental}).

With the calculated phonon frequencies, the linear Gr\"{u}neisen parameter $\gamma_{ab}$ is calculated by varying the $a$ and $b$ dimensions of the unit cell by $0.005\%$ with fixed $c$ dimension,
\begin{eqnarray}\label{gammaa}
\gamma_{ab}=\left(-\partial \ln\omega/\partial \ln a\right)_c
\end{eqnarray}
and the linear Gr\"{u}neisen parameter $\gamma_{c}$ is calculated by varying the $c$ dimension of the unit cell by $0.005\%$ with fixed $a$ and $b$ dimensions,
\begin{eqnarray}\label{gammac}
\gamma_{c}=\left(-\partial \ln\omega/\partial \ln c\right)_{ab}.
\end{eqnarray}
We will show later how these two quantities determine the coefficients of linear thermal expansion $\alpha_a = \partial \ln a/\partial T$ and $\alpha_c =\partial \ln c/\partial T$.

\section{Ground-state properties of Ag$_3$Co(CN)$_6$}\label{groundstate}

\subsection{Crystal structure}

The details ground-state structures of Ag$_3$Co(CN)$_6$ optimised using GGA, with and without the dispersive interaction, and using LDA are reported in Table~\ref{tab:groundstate}, where they are compared to the experimental values~\cite{Conterio2008}. It is clear that, without the dispersive interaction, the calculated ground-state structure is wrong. Inclusion of the dispersive interaction results in the correct structure with small deviations from experiment.

It is worth remarking on the role the Ag$\ldots$Ag dispersive interaction has on the structure. The dispersive interaction is a weak attractive interaction, which opposing the repulsive Coulomb interaction, Thus the effect of the dispersive interaction is to reduce the overall Ag$\ldots$Ag interaction. On this basis, addition of the dispersive interaction to the GGA model enables the structure to relax with a shorter Ag$\ldots$Ag distance and hence a smaller value of the $a$ lattice parameter, as see in the results in Table \ref{tab:groundstate}. On the other hand, the well-known tendency of LDA to overbind already results in a shorter Ag$\ldots$Ag distance.

We can quantify this point. The DFT calculations give an approximate value for the charge of the Ag cation of $+0.65|e|$\cite{Segall1996}, where $e$ is the electronic charge. Calculation of the Ag$\ldots$Ag forces due to the Coulomb and dispersive interactions (taking $f_\mathrm{damp}=1$ in Eq.~\ref{g06}) over the range of distances 3.3--3.5~\AA\ shows that the dispersive interaction reduces the net force between neighbouring Ag ions by nearly a factor of 2.

\subsection{Elasticity}

The GGA+D computed elastic compliances are given in Table~\ref{tab:compliance}. The linear compressibilities along the $a$($b$) and $c$ crystal axes were calculated using the elastic compliances as
\begin{eqnarray}\label{linearcompressibility}
\beta _{ab}  = -\partial \ln a/\partial p  =s_{11}  + s_{12}  + s_{13}
\end{eqnarray}
and
\begin{eqnarray}\label{linearcompressibility2}
\beta _c  =  -\partial \ln c/\partial p = 2s_{13}  + s_{33},
\end{eqnarray}
respectively. The volume compressibility was calculated as the sum
\begin{eqnarray}\label{linearcompressibility3}
\beta = -\partial \ln V/\partial p =2\beta_{ab}  + \beta_c
\end{eqnarray}
The linear elastic moduli $B_{ab}$ along $a$ and $b$ axes as well as $B_c$ along $c$ axis are the inverse of the $\beta_{ab}$ and $\beta_c$, respectively. Their relations with the elastic constants are given in the supplemental material~\cite{supplemental}.

As shown in Table~\ref{tab:compliance}, the GGA+D calculated $s_{33}$ and $s_{13}$ have almost the same magnitude but with opposite sign, showing that the $c$ dimension would response equivalently to a stress acting on the $a$ or $b$ dimension and a tension directly acting on the $c$ dimension. This shows the  effectiveness of the hinging mechanism in the material. In comparison, the small value of $s_{12}$ shows that the change in dimension $a$ (or $b$) is barely correlated to the change in $b$ (or $a$) dimension.

Negative values of $\beta_c$ and $B_c$ correspond to the NLC of the material, namely the material will \textit{elongate} in the $c$ dimension under hydrostatic compression. The bulk modulus and its first derivative were calculated as $B=15.8(8)$~GPa and $B^\prime=-4.9(8)$, respectively~\cite{supplemental}. Using the 3rd-order Birch-Murnaghan (BM) equation of state (EoS)~\cite{Birch1947} to fit to the calculated isotherm data from 0 to 0.6~GPa also results in a negative value of $B^\prime$ of $-3(2)$. These results predict that the material will have pressure-induced softening~\cite{Fangzeolite2013,Fangexp2013,Fangexpression2014} at low pressures.

\begin{table}[t]
\setlength{\tabcolsep}{4pt}
\caption{\label{tab:compliance} Calculated compliances at different pressures for the ambient phase of Ag$_3$Co(CN)$_6$ obtained from calculating the change of energy corresponding to a set of given strains $\varepsilon _{ij}$ generated according to the trigonal symmetry. Results were obtained using GGA+D, and the LDA results are from Ref.~\onlinecite{Hermet2013}.The compliances of a trigonal phase have the symmetry~\cite{Nye1985} $s_{ij}=s_{ji}, s_{22}=s_{11}, s_{55}=s_{44}, s_{23}=s_{13}, s_{24}=-s_{14}, s_{66}=2(s_{11}-s_{12})$. The corresponding elastic constants and elastic moduli are given in the Supplemental Material~\cite{supplemental}. The linear compressibility $\beta_{ab}$ and $\beta_c$ as well as the volume compressibility $\beta$ are calculated from the compliances using Eqs~\eqref{linearcompressibility} to~\eqref{linearcompressibility3}.}
\centering
\begin{tabular}{ccccc}
\hline Compliance (TPa$^{-1}$) & 0.0 GPa & 0.04 GPa & 0.1 GPa & LDA \\
\hline
$s_{11}$ & 61(3)  & 62(3)  & 64(4)  & 85 \\
$s_{33}$ & 22(1)  & 21.4(9)   & 23(2)  & 16\\
$s_{44}$ & 38.5(9)   & 37.7(7)   & 44(3)  & 73 \\
$s_{12}$ & 2(1)   & 1(1)   & 3(2)   & $-22$ \\
$s_{13}$ & $-21(1)$   & $-21(1)$   & $-23(2)$   & $-17$ \\
$s_{14}$ & $15(1)$ & $15(1)$ & $17(2)$ & $-41$ \\
$\beta_{c}$    & $-21(2)$ & $-21(2)$ & $-23(4)$ & $-19$ \\
$\beta_{ab}$   & $42(4)$  & 42(4)    & 44(5)    & 45    \\
$\beta$        & $63(6)$  & 63(6)   & 65(8)    & 72    \\
\hline
\end{tabular}
\end{table}

The calculated bulk modulus at 0 K, as the inverse of $\beta$ in Eq.~\ref{linearcompressibility3}, is 15.8(8)~GPa which is significantly larger than the experimental value of $B=6.5(3)$~GPa at 300~K~\cite{GoodwinAgCoCNnlc2008}. This apparent overestimation of the calculation may actually be due to a considerable softening of the material on heating, as will be discussed later in Section~\ref{softening}. The same idea can be used to explain the apparent large underestimation of the compressibilities: the calculated values $\beta_{ab}=42(4)$~TPa$^{-1}$ and $\beta_c=-21(2)$~TPa$^{-1}$ are much lower than the experimental values of $\beta_{ab}=115(8)$~TPa$^{-1}$ and $\beta_c=-79(9)$~TPa$^{-1}$ at 300~K~\cite{GoodwinAgCoCNnlc2008}.

\begin{figure}[t]
\begin{center}
\includegraphics[width=0.5\textwidth]{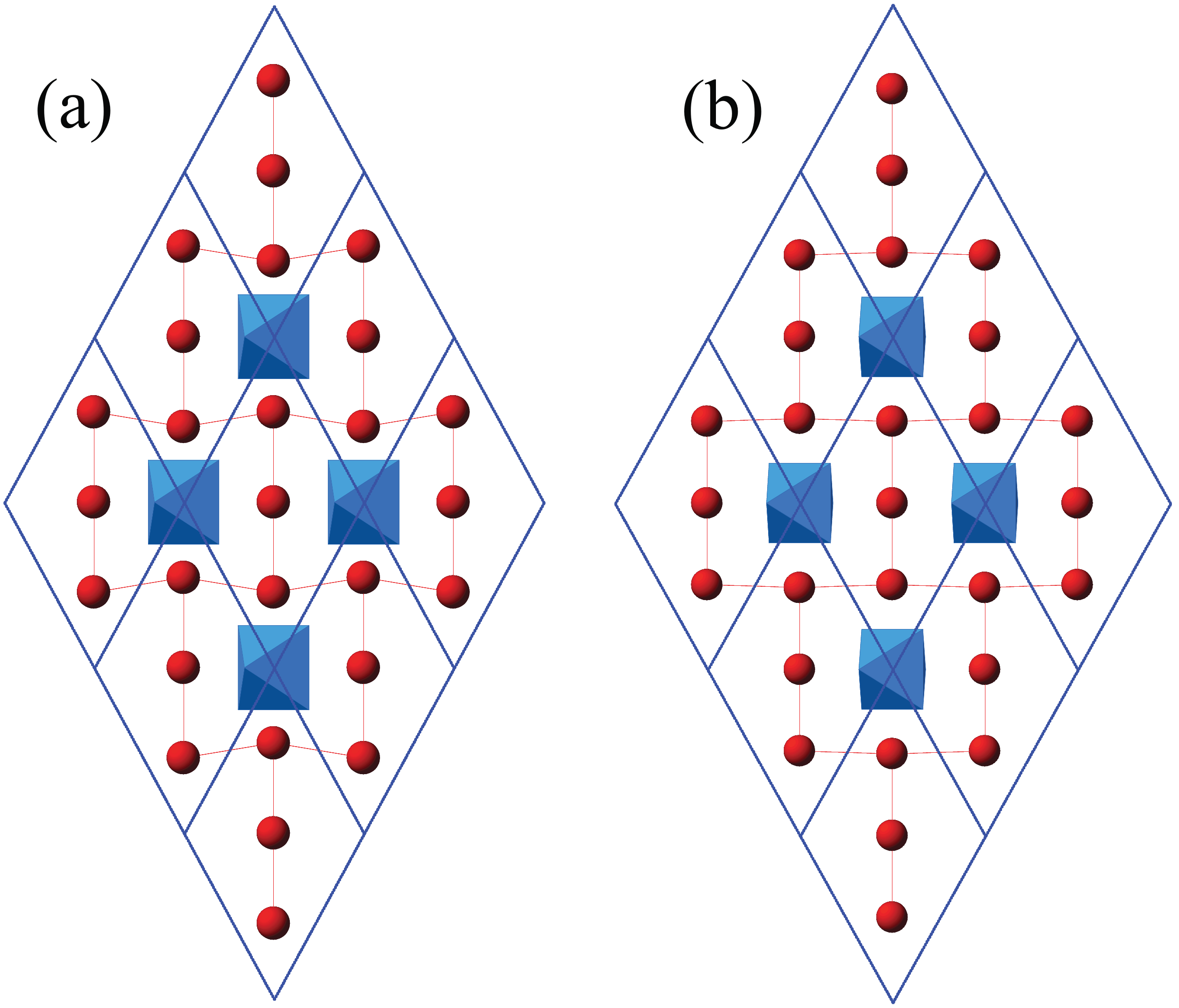}
\end{center}
\caption{\label{fig:phaseII} Structures of the high-pressure phase of Ag$_3$Co(CN)$_6$ (space group $C2/m$) optimised using (a) GGA+D , and (b) either GGA or LDA without a correction for the dispersion energy. The experimentally observed interdigitated structure, characterised by the indented Ag atoms, can be seen only when dispersion corrections are used.}
\end{figure}

\section{High-pressure phase of Ag$_3$Co(CN)$_6$}

\subsection{Crystal structure of the high-pressure phase}
Ag$_3$Co(CN)$_6$ undergoes a structural phase transition at $0.19$ GPa to a monoclinic phase~\cite{GoodwinAgCoCNnlc2008} and denoted as Phase-II. The phase transition involves displacements of Ag atoms in alternative rows, which cause the high-pressure phase to possess an interdigitated structure as seen by viewing down the $\left[0,\,0,\,1\right]$ direction. This is indicated in Fig.~\ref{fig:phaseII}(a) by the indented Ag atoms.

\begin{figure}[t]
\begin{center}
\includegraphics[width=0.48\textwidth]{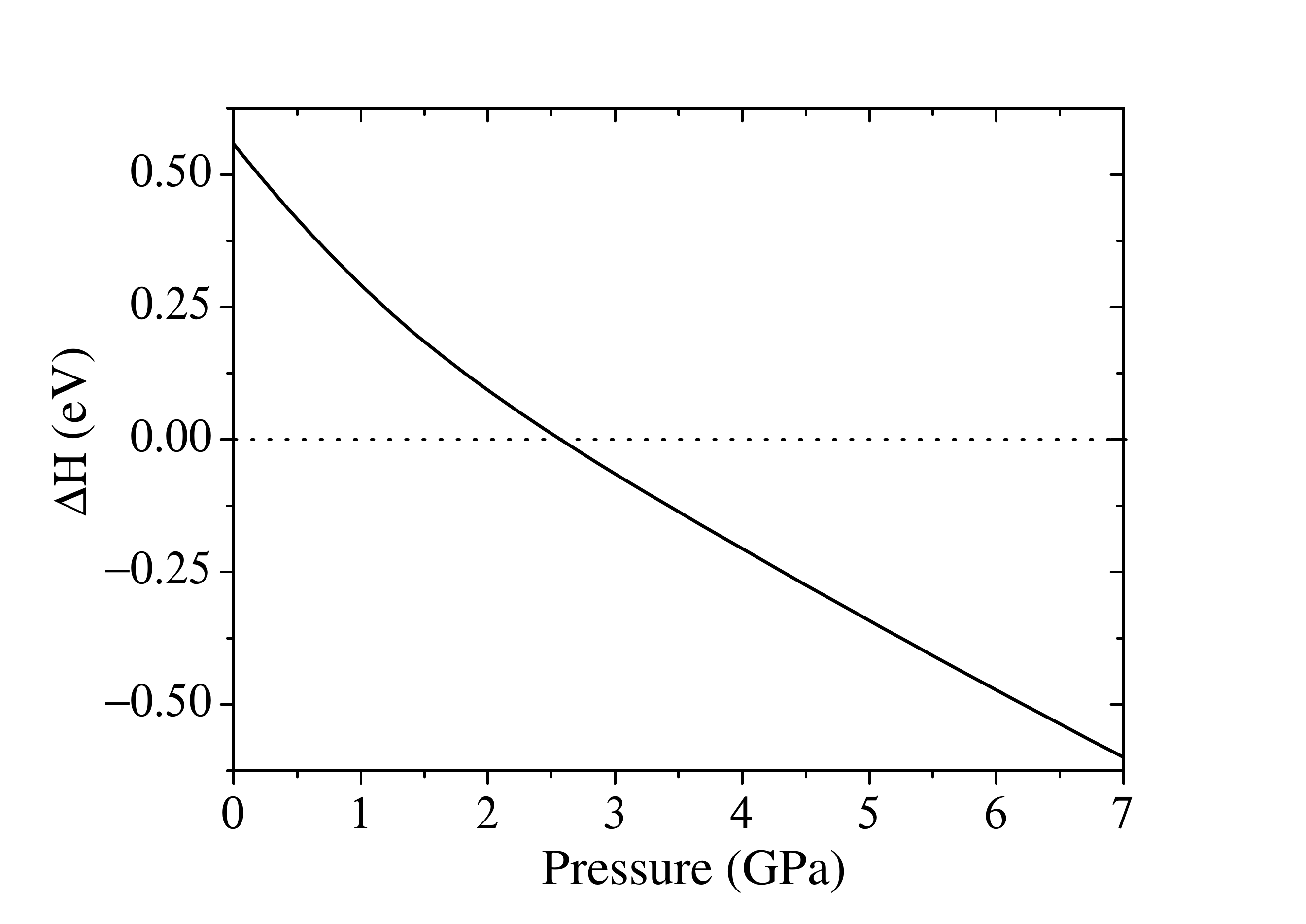}
\end{center}
\caption{\label{fig:phasetransition2} The calculated enthalpy of the high-pressure phase relative to that of the ambient phase using GGA+D. An overestimated phase-transition pressure of $2.5$~GPa is predicted.}
\end{figure}

Our calculations show that neither LDA or GGA without the dispersive interaction can give the correct optimised high-pressure phase with the interdigitated structure~\cite{GoodwinAgCoCNnlc2008}, as shown by Fig.~\ref{fig:phaseII}(b). It is only by including the dispersive interaction in the GGA+D calculation that the interdigitated structure of the high-pressure phase can be reproduced, as shown in Fig.~\ref{fig:phaseII}(a).

Fig.~\ref{fig:phasetransition2} shows the difference in enthalpy between the two phases as calculated using the GGA+D method. The predicted phase-transition pressure of about 2.5 GPa overestimates the experimental value of 0.19 GPa~\cite{GoodwinAgCoCNnlc2008}. Although this appears to be a large discrepancy, it is magnified by the fact that the experimental transition pressure is so low. Phase transition pressures are hard to calculate; we attribute the discrepancy to an accumulation of small errors associated with a number of approximations in the DFT method and the dispersion correction. The calculated relative change of the cell volume at the phase transition is $11\%$, smaller than the experimental value of $16\%$~\cite{GoodwinAgCoCNnlc2008}. Table~\ref{tab:highpphase} compares the optimised structure with the $C2/m$ space group in GGA+D with the experiment values at 0.23 GPa.

\begin{table}[t]
\caption{\label{tab:highpphase} Comparison of optimised (GGA+D2) and experimental \cite{GoodwinAgCoCNnlc2008} crystal structures of the high-pressure phase (space group $C2/m$) at a pressure of 0.23 GPa. $\Delta_\mathrm{GGA+D}$ represents the differences between the two. $V$ is the volume of one formula unit (note that there are 2 formula units in the unit cell). The fractional coordinates of Ag1 are $(1/2, 0, 1/2)$. }
\begin{tabular}{@{\extracolsep{10pt}}c|ccc}
\hline  & GGA+D & Experiment & $\Delta_\mathrm{GGA+D}$ \\
\hline
$a$ (\AA) & $6.485$  & 6.693  & $-3.1$\%\\
$b$ (\AA) & $11.144$ & 11.539 & $-3.4$\% \\
$c$ (\AA) & $6.658$  & 6.566  & $+1.4$\%\\
$\beta$ ($^\circ$)     & $101.84$ & $101.48$ & $+0.36$ \\
$V$ (\AA$^3$)& 235.6    & 248.5   & $+5.2$\% \\
C1$_x$       & 0.790& 0.825  & $-0.035$  \\
C1$_z$       & 0.163& 0.182  & $-0.019$   \\
N1$_x$       & 0.664& 0.715  & $-0.051$   \\
N1$_z$       & 0.264& 0.302  & $-0.038$   \\
C2$_x$       & 0.145& 0.163  & $-0.019$   \\
C2$_y$       & 0.123& 0.119  & $+0.004$   \\
C2$_z$       & 0.177& 0.157  & $+0.0209$   \\
N2$_x$       & 0.241& 0.258  & $-0.017$   \\
N2$_y$       & 0.197& 0.185  & $+0.012$   \\
N2$_z$       & 0.280& 0.259  & $+0.021$   \\
Ag2$_y$      & 0.243& 0.240  & $+0.002$   \\
C1--N1 (\AA) & 1.161 & 1.183 & $-1.8$\%\\
C2--N2 (\AA) & 1.170 & 1.126 & $+3.9$\%\\
Ag1--N1 (\AA)&2.069 & 2.123 & $-2.5$\%\\
Ag2--N2 (\AA)&2.097 & 2.199 & $-4.6$\%\\
Co--C1 (\AA) &1.907 & 1.830 & $+4.2$\%\\
Co--C2 (\AA) &1.922 & 1.924 & $-0.1$\%\\
Ag--Ag(1) (\AA)&2.868&2.996 & $-4.3$\%\\
Ag--Ag(2) (\AA)&5.407&5.548 & $-2.5$\%\\
\hline
\end{tabular}
\end{table}

Originally, it was found~\cite{GoodwinAgCoCNnlc2008} that the high-pressure phase of the material has a space group of $C2/m$. However, recently, it was proposed~\cite{Hermet2013} that the high-pressure phase should have the lower symmetry of space group $Cm$, because a structure with this symmetry can be obtained as a subgroup of the space group of the ambient-pressure phase, $P\bar{3}1m$, whereas a structure with space group $C2/m$ cannot. Our calculations indicate that the optimised structures starting from both space groups $C2/m$ and $Cm$ have exactly the same enthalpy up to a pressure of 7~GPa (the highest we examined), with relaxed structures that differ only by a small origin offset. We conclude that the structure of the high-pressure phase has the originally-proposed $C2/m$ structure.

\begin{figure}[t]
\begin{center}
\includegraphics[width=0.48\textwidth]{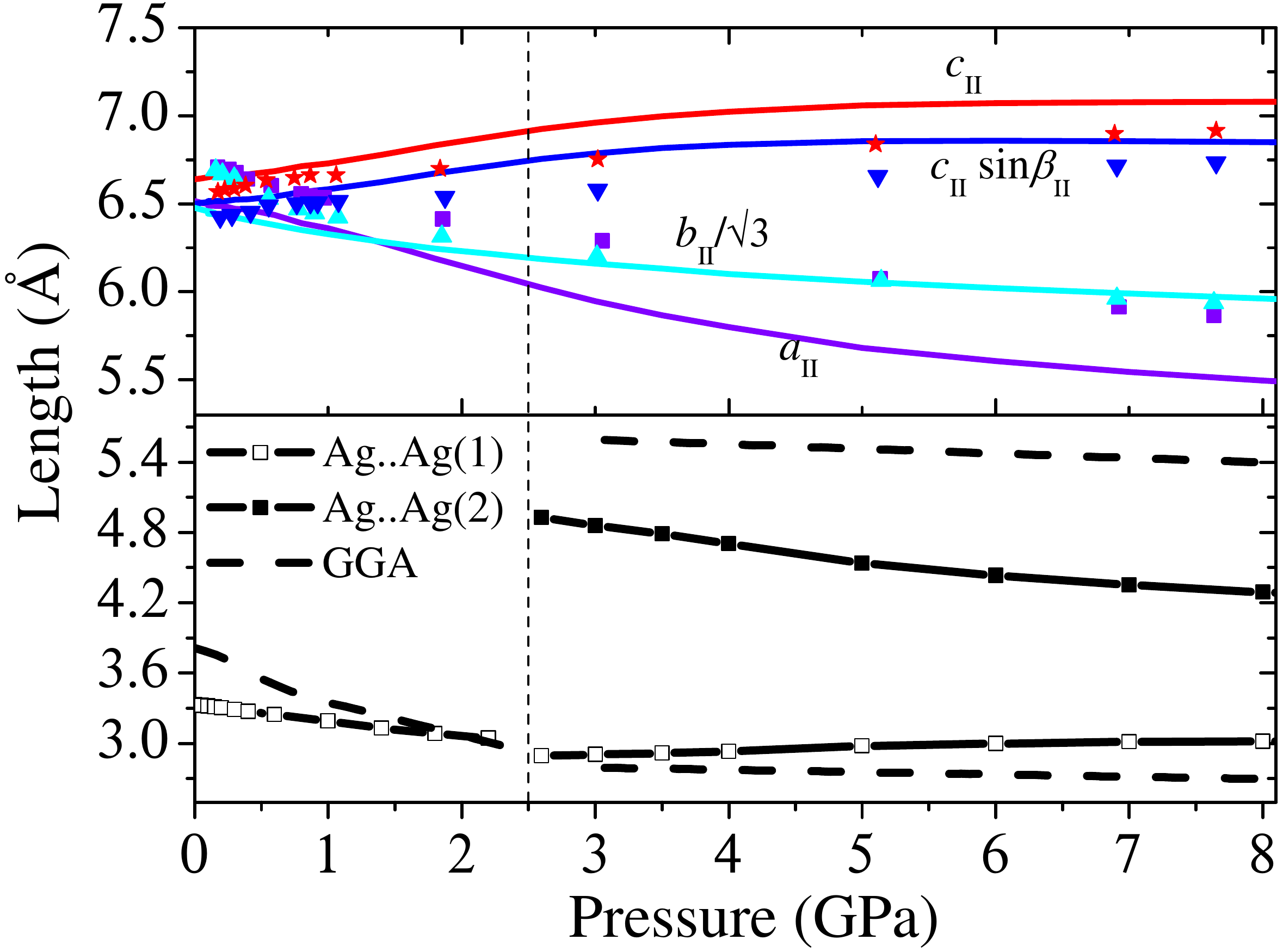}
\end{center}
\caption{\label{fig:phaseIIparameters} Upper panel: various lattice parameters of the monoclinic phase of Ag$_3$Co(CN)$_6$ at different pressures using GGA+D. The calculated (solid lines) and the experimental (symbol) values of each parameter are in the same colour. Lower panel: the calculated nearest Ag$\ldots$Ag distance in two phases. In the high-pressure phase, the GGA+D result (symbol line) shows the correct trend of two types of Ag$\ldots$Ag distances changing with compression, while the GGA/LDA (dashed line) result does not.}
\end{figure}

\subsection{Elasticity}

A fit of the 3rd-order BM EoS to the calculated isotherm of the high-pressure phase yields $B=17(6)$~GPa and $B^\prime=17(7)$; experimental values are $B=11.8(7)$~GPa and $B^\prime=13(1)$, respectively~\cite{GoodwinAgCoCNnlc2008}. Thus, unlike the ambient-pressure phase, which has pressure-induced softening at low pressures, the high-pressure phase of the material quickly becomes harder under compression.

The calculated change of lattice parameters of the high-pressure monoclinic phase-II are presented in Fig.~\ref{fig:phaseIIparameters}, and compared to the experimental values. The agreement between the two are good with the largest relative deviation below $10\%$. By fitting to a 3rd-order polynomial of pressure $\left(p-p_\mathrm{c}\right)$ with the phase-transition pressure $p_\mathrm{c}=2.5$~GPa, the linear compressibilities of $a_\mathrm{II}$, $b_\mathrm{II}$ and $c_\mathrm{II}$ were obtained at different pressures. Their averaged values over $2.5$--$8.0$~GPa are $19(1)$, $6.9(4)$ and $-4.1(3)$~TPa$^{-1}$, respectively. These values are in good agreement with experimental values~\cite{GoodwinAgCoCNnlc2008} of $15.9(9)$, $9.6(5)$ and $-5.3(3)$~TPa$^{-1}$.

As pointed out in Ref.~\onlinecite{GoodwinAgCoCNnlc2008}, the relatively small compressibility along $b_\mathrm{II}$ is due to the interdigitation in the high-pressure phase. Upon compression, the structure becomes more indented (Fig.~\ref{fig:phaseII}(b)), resulting in the Ag$\ldots$Ag(1) distance between the indented Ag atom and its nearest neighbour increases with pressure, while the Ag$\ldots$Ag(2) distance between the two indented Ag atoms at the opposite sites decreases. This behaviour of the Ag$\ldots$Ag distances under pressure is seen in the GGA+D calculated results shown in the lower panel of Fig.~\ref{fig:phaseIIparameters}.

\begin{table}[t]
\setlength{\tabcolsep}{3pt}
\caption{\label{tab:raman} The calculated Raman and infrared spectrums (in THz) of Ag$_3$Co(CN)$_6$ using DFPT+D compared to the experimental values at 80 K (Raman)~\cite{Rao2011} and 295 K (Infrared)~\cite{Hermet2013}. $\Delta_\mathrm{DFPT+D}$ is the deviation of the DFPT+D calculated frequencies compared to the experiment (in THz). The first derivative of the frequency with respect to pressure is in unit of THz/GPa.}
\centering
\begin{tabular}{ccccc}
\hline Raman & $\omega_\mathrm{DFPT+D}$ & $\Delta_\mathrm{DFPT+D}$ & $\left(\partial \omega /\partial p\right)_\mathrm{Exp.}$~\cite{Rao2011} & $\left(\partial \omega /\partial p\right)_\mathrm{DFPT+D}$ \\
\hline
$2.6$  & 2.9  & 0.3 & 0.3   & 0.4  \\
$4.2$  & 4.3  & 0.1 & 0.6   & 0.4  \\
$4.9$  & 5.0  & 0.1 & 0.3   & 0.6  \\
$9.7$  & 9.8  & 0.1 & $-0.3\footnotemark[1]$&$-0.04$ \\
$14.2$ & 13.8 & $-0.4$ & 0.1  &$-0.006$ \\
$14.2$ & 13.9 & $-0.3$ & 0.1  &$-0.04$ \\
$15.6$ & 16.1 & 0.5    & 0.7   & 0.06  \\
$15.6$ & 16.1 & 0.5    & 0.7   & 0.2  \\
$65.5$ & 65.2 & $-0.3$ & 0.2  & 0.1 \\
$66.1$ & 66.0 & $-0.1$ & 0.3   & 0.1 \\
\hline Infrared & $\omega_\mathrm{DFPT+D}$ & $\Delta_\mathrm{DFPT+D}$ & $\left(\partial \omega /\partial p\right)_\mathrm{Exp.}$~\cite{Hermet2013} & $\left(\partial \omega /\partial p\right)_\mathrm{DFPT+D}$  \\
\hline
1.2  & 1.4   & 0.2 & --     & 0.2       \\
1.4  & 1.5   & 0.1  & --     & $0.1$   \\
1.6  & 2.2   & 0.6 & --     & $-0.3$   \\
4.0  & 4.2   & 0.2  & $-0.2$ & $0.01$   \\
5.3  & 5.6   & 0.3  & --     & $-0.4$   \\
5.5  & 5.7   & 0.2  & --     & $-0.2$  \\
8.0  & 8.6   & 0.6  & --     & 0.3    \\
8.0  & 8.8   & 0.8 & --     & 0.2       \\
13.0 & 12.8  & $-0.2$  & --     & $-0.03$   \\
14.5 & 14.8  & 0.3  & $-0.02$ & $0.01$  \\
14.8 & 14.9  & 0.1  & 0.03    & 0.01    \\
17.6 & 17.8  & 0.2  & --     & 0.3     \\
--   & 18.0  & --   & --     & 0.2     \\
--   & 65.1  & --   & --     & $0.1$ \\
--   & 65.2  & --   & --     & 0.1    \\
\hline
\end{tabular}
\footnotetext[1]{From non-hydrostatic experiment~\cite{Rao2011}.}
\end{table}

\begin{figure*}[t]
\begin{center}
\includegraphics[width=0.97\textwidth]{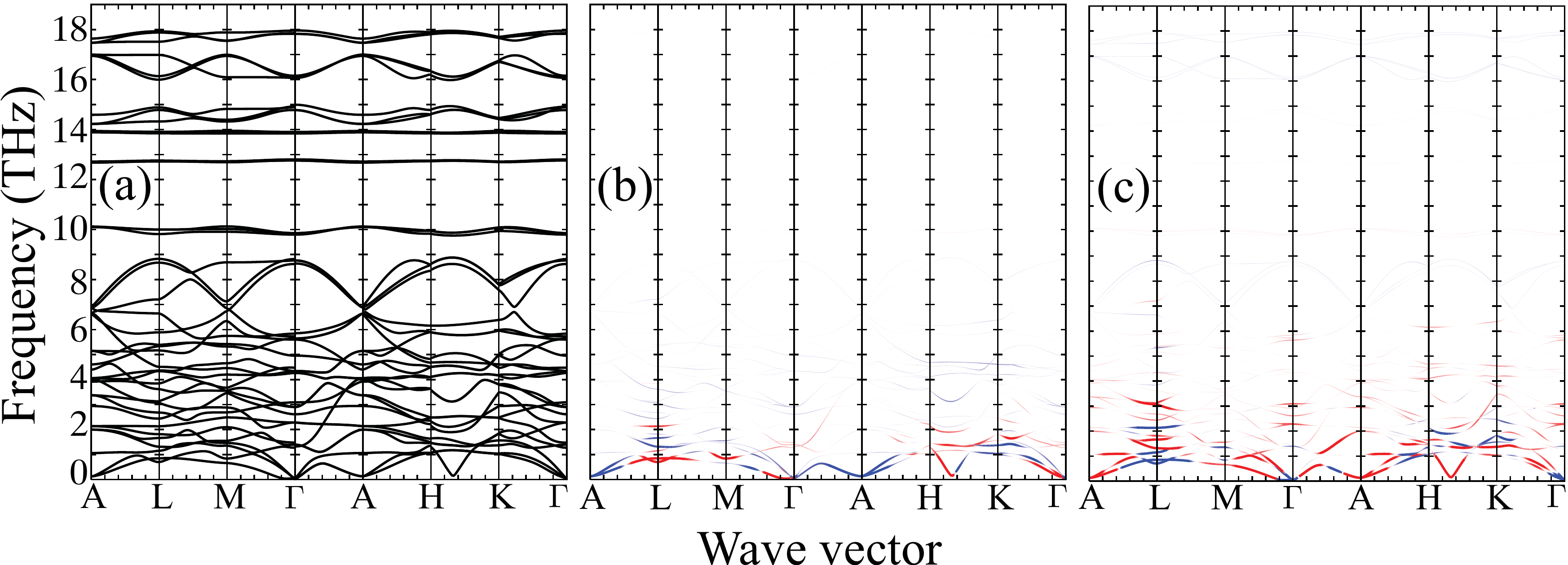}
\end{center}
\caption{\label{fig:phonon} (a) DFPT+D calculated phonon dispersion curves along the high-symmetry directions in the Brillouin zone. (b) and (c) are dispersion curves coloured according to the values of linear Gr\"{u}neisen parameters along the $a$($b$) axes ($\gamma_{ab}$) and $c$ axis ($\gamma_c$), respectively, with values $\leq-20$ in red gradually passing to values $\geq+20$ in blue.}
\end{figure*}

\section{Lattice dynamics calculations}

The phonon calculations were performed using the DFPT+D method as discussed in Section~\ref{section:DFTPD2methods}. Table~\ref{tab:raman} shows that the calculated Raman and infrared spectra are in good agreement with the experiment~\cite{Rao2011,Hermet2013}. The phonon dispersion curves along the high-symmetry directions in the Brillouin zone for frequencies up to 18~THz are presented in Fig.~\ref{fig:phonon}(a).

We have studied the eigenvectors of different vibrational modes as shown by the animations in the Supplemental Materials~\cite{supplemental}. We found that the infrared-active modes at $1.4$--$1.5$~THz showing negative linear Gr\"{u}neisen parameters $\gamma_{ab}$ and positive linear Gr\"{u}neisen parameters $\gamma_c$ correspond to the rotation of Ag-triangle pairs against each other in the Kagome sheet about their shared apex. The Raman-active mode at $2.9$~THz, having positive $\gamma_{ab}$ and negative $\gamma_c$, corresponds to the rotations of  CoC$_6$ octahedra that pulls the connected layers of Ag atoms along the $c$ axis closer together. The Raman-active modes at $4.3$ and $5.0$~THz correspond to similar type of vibrations but with CoC$_6$ octahedra deforming, and these also show positive $\gamma_{ab}$ and negative $\gamma_c$.

\begin{figure}[t]
\begin{center}
\includegraphics[width=0.48\textwidth]{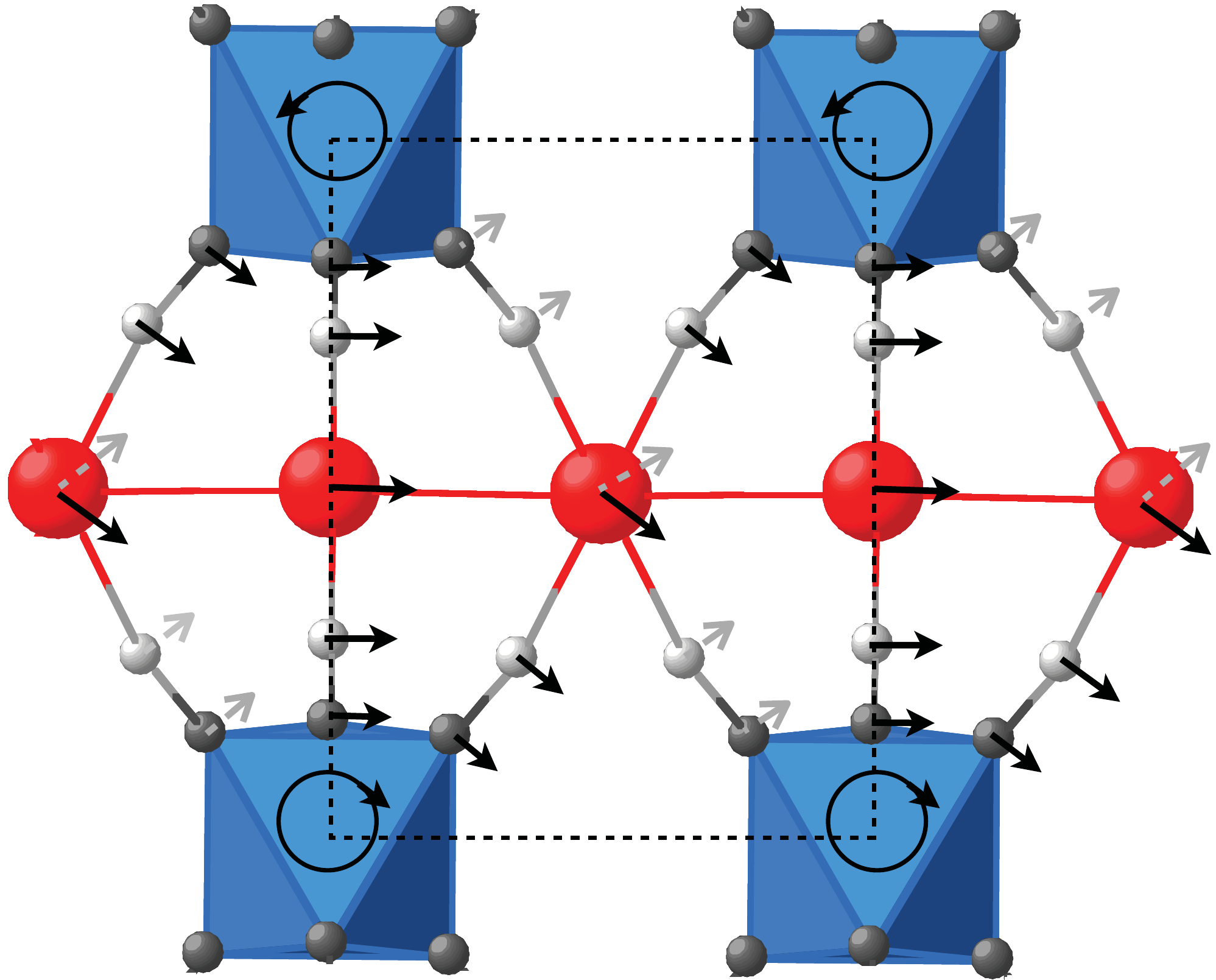}
\end{center}
\caption{\label{fig:Amode1} Vibration corresponds to the first mode at point A $\left(0,\,0,\,1/2\right)$ from its eigenvector looking down the $\left[1,\,0,\,0\right]$ direction. Each Ag atom (red) is connected to two [Co(CN)$_6$] octahedra (blue) in the upper and lower layers via the Co--CN--Ag--NC--Co linkages. Arrows show the transverse motion of the nearly-rigid bridging group CN--Ag--NC resulted from the concerted rotation of the octahedra. The dashed square shows the unit cell.}
\end{figure}

The dispersion curves are also shown in Fig.~\ref{fig:phonon}(b) and (c) with colours that reflect the calculated values of $\gamma_{ab}$ and $\gamma_c$ as given by Eqs~\ref{gammaa} and~\ref{gammac}, respectively. One can see that it is almost the same set of low-frequency modes that contribute to the PTE along the $a$($b$) axes and NTE along the $c$ axis, i.e. their values of $\gamma_{ab}$ and $\gamma_c$ show similar magnitudes but are opposite in sign. This is directly related to the hinging structure in the material where any level of expansion in the $a$($b$) axes would transfer into a similar level of contraction in the $c$ axis via the Co--CN--Ag--NC--Co linkage. Modes around the wave vector A $\left(0,\,0,\,1/2\right)$ and around the middle point along the H$\left(-1/3,\,2/3,\,1/2\right)$$\rightarrow$K$\left(-1/3,\,2/3,\,0\right)$ direction have the lowest frequencies ($< 1.0$ THz) and hence have the most extreme values of Gr\"{u}neisen parameters, The first two degenerate modes at A correspond to concerted rotations of rigid Co(CN)$_6$ octahedra together with the nearly-rigid CN--Ag--NC linkages moving sideways~\cite{supplemental}, as shown by its eigenvector in Fig.~\ref{fig:Amode1}. The first mode at the middle point $\left(-1/3,\,2/3,\,1/4\right)$ along the H$\rightarrow$K direction corresponds to the Ag atoms vibrating along the $c$ axis, producing a transverse wave passing through each Kagome sheet~\cite{supplemental}.

\begin{figure*}[t]
\begin{center}
\includegraphics[width=0.97\textwidth]{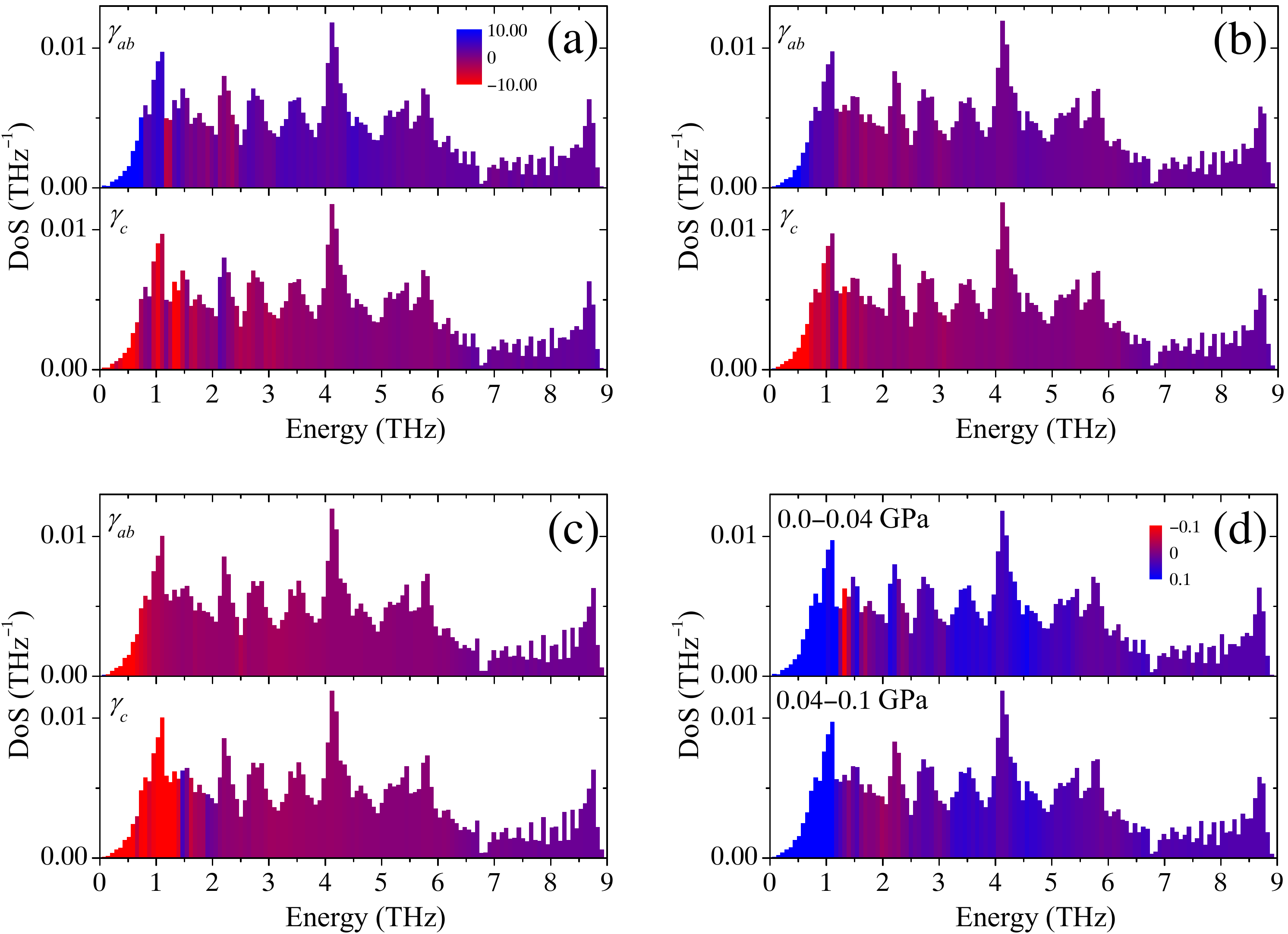}
\end{center}
\caption{\label{fig:dosgp} Calculated DoS of modes with frequencies $\leq 9$ THz. At pressures (a) 0.0 GPa, (b) 0.04 GPa and (c) 0.1 GPa, the DoS in the upper panel is coloured according to the averaged value of $\gamma_{ab}$ and the DoS in the lower panel is coloured according to the averaged value of $\gamma_{c}$ around each energy. Values $\leq -10$ are in red and $\geq 10$ are in blue. $\gamma_{ab}$ of the low-frequency modes, especially the modes below 1.0 THz, decrease largely upon compression and even change their signs at 0.1 GPa as indicated by the change of colour from blue to red. (d) Coloured DoS according to the average value of relative frequency change with pressure (in GPa$^{-1}$) around each energy bin. The upper panel shows the frequency change from 0.0 to 0.04 GPa and the lower panel shows that from 0.04 to 0.1 GPa. Stiffened phonons ($\partial \ln \omega/\partial p \geq +0.1$) are in blue and softened phonons ($\partial \ln \omega/\partial p \leq -0.1$) are in red.}
\end{figure*}

The picture shown in Fig.~\ref{fig:phonon} is reflected in plots of the vibrational densities of states (DoS), which are shown in Fig.~\ref{fig:dosgp}. These were calculated from the full set of DFPT+D vibrations computed on a $25\times25\times25$ grid (corresponding to a total of 1470 wave vectors in the Brillouin zone). Plots of the DoS are plotted for three pressures and coloured according to the averaged value of $\gamma_{ab}$ and $\gamma_c$ of the modes around each energy. The plots for vibrations at ambient pressure (Fig.~\ref{fig:dosgp}(a)) show that the same low-frequency modes contribute positively to $\gamma_{ab}$ and negatively to $\gamma_c$. This situation changes under pressure, as we will now discuss.

\section{Effect of compression on thermal expansion}

\subsection{Increase of linear thermal expansion on compression}

From the calculated Gr\"{u}neisen parameters and the compliances given in Table~\ref{tab:compliance}, the linear coefficients of thermal expansion of Ag$_3$Co(CN)$_6$ along the $a$($b$) and the $c$ axes were calculated within the quasi-harmonic approximation as~\cite{Barron1980}
\begin{eqnarray}\label{linearcteab}
\alpha _{ab}  &=& \frac{1}{\Omega }\sum\limits_{s,\textbf{k}} {\left\{ {c_{s,\textbf{k}} \left[ {\frac{{\left( {s_{11}  + s_{12} } \right)\gamma _{ab}(s,\textbf{k}) }}{2} + s_{13} \gamma _c(s,\textbf{k}) } \right]} \right\}} \nonumber \\
&=& \frac{1}{\Omega }\left[ {\frac{{\left( {s_{11}  + s_{12} } \right)\overline \gamma _{ab} }}{2} + s_{13} \overline \gamma _c } \right]
\end{eqnarray}
and
\begin{eqnarray}\label{linearctec}
\alpha _c  &=& \frac{1}{\Omega }\sum\limits_{s,\textbf{k}} {\left\{ {c_{s,\textbf{k}} \left[ {s_{13} \gamma _{ab}(s,\textbf{k})  + s_{33} \gamma _c(s,\textbf{k}) } \right]} \right\}} \nonumber \\
&=& \frac{1}{\Omega }\left[ {s_{13} \overline \gamma _{ab}  + s_{33} \overline \gamma _c } \right],
\end{eqnarray}
respectively, where
\begin{eqnarray}\label{specificheat}
c_{s,\textbf{k}}  = \hbar \omega _{s,\textbf{k}} \frac{{\partial n_{s,\textbf{k}} }}{{\partial T}}
\end{eqnarray}
is the contribution of the normal-mode $\left\{s,\textbf{k}\right\}$ to the specific heat with $n_{s,\textbf{k}}  = \left[ {\exp \left( {\hbar \omega _{s,\textbf{k}} /k_BT } \right) - 1} \right]^{ - 1}$, and $\Omega$ is the volume of the unit cell. The overall Gr\"{u}neisen parameters are defined as
\begin{eqnarray}\label{overallgamma}
\overline \gamma_{ab}&=&\sum\limits_{s,\textbf{k}} {c_{s,\textbf{k}} \gamma_{ab}(s,\textbf{k})} \nonumber \\
\overline \gamma_{c}&=&\sum\limits_{s,\textbf{k}}  {c_{s,\textbf{k}} \gamma_{c}(s,\textbf{k})}.
\end{eqnarray}
The volume CTE is calculated as
\begin{eqnarray}\label{linearctev}
\alpha _V  = 2\alpha _{ab}  + \alpha _c.
\end{eqnarray}

\begin{figure}[t]
\begin{center}
\includegraphics[width=0.48\textwidth]{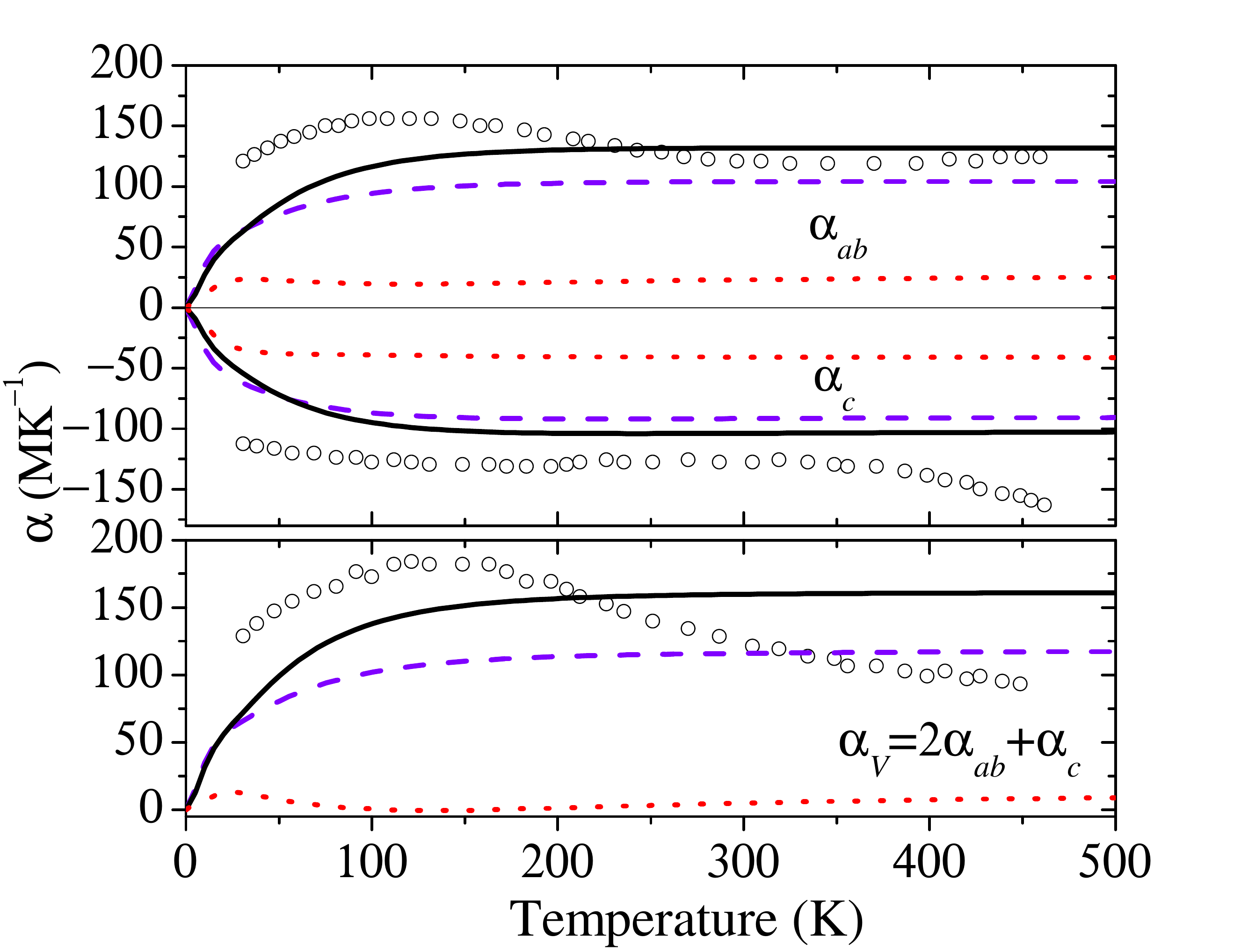}
\end{center}
\caption{\label{fig:cote} DFPT+D calculated coefficients of thermal expansion at different temperatures for pressures of 0.0 (solid line), 0.04 (dashed line) and 0.1 GPa (dotted line) using quasi-harmonic approximation, compared to the experiment at ambient pressure (in open circle)~\cite{GoodwinAgCoCN2008}. A plot from GGA calculated phonons based on the correct structure optimised using the GGA+D method is given in the Supplemental Material~\cite{supplemental} for comparison.}
\end{figure}

The calculated values of $\alpha_{ab}$ and $\alpha_c$ at different temperatures and pressures are shown in Fig.~\ref{fig:cote}. The averaged values of $\alpha_{ab}$ and $\alpha_c$ over 50--500 K are $+127$~MK$^{-1}$ and $-101$~MK$^{-1}$, respectively. These exceptionally large values are in reasonable agreement with the experimental values~\cite{GoodwinAgCoCN2008} of $\alpha_{ab}=+135$~MK$^{-1}$ and $\alpha_c=-131$~MK$^{-1}$. The hinging mechanism of the material as discussed previously results in similar magnitude of the PTE along the $a$($b$) axes and the NTE in the $c$ axis.

In addition to reproducing the experimentally-observed~\cite{GoodwinAgCoCN2008} colossal PTE and NTE of Ag$_3$Co(CN)$_6$, an interesting finding from Fig.~\ref{fig:cote} is that $\partial \alpha_c / \partial p > 0$, that is $\alpha_{c}$, which has a negative value, becomes less negative on compression. This is opposite to the usual behaviour that $\partial \alpha/ \partial p<0$ as found in most PTE materials such as metals, metal oxides and alkali halides~\cite{Fangmetal2010,Zhang2007,Song2012,Sun2013} and also in many isotropic NTE materials~\cite{Chapman2005,Fangmd2013,Fangzeolite2013,Fangmodel2014}.

According to the standard thermodynamic relation~\cite{Fangexp2013}
\begin{eqnarray}\label{linearwarmhardening}
\left( {\frac{{\partial B_c }}{{\partial T}}} \right)_p  = B_c^2 \left( {\frac{{\partial \alpha _c }}{{\partial p}}} \right)_T,
\end{eqnarray}
a positive value of $\partial \alpha_c / \partial p$ means a corresponding positive value of $\partial B_c / \partial T$. If $B_c$ were positive as would usually be the case, this would give the unusual property of the material becoming harder at higher temperature~\cite{Fangmodel2014}, but in this case $B_c$, as the inverse of $\beta_c$ is negative (see Table~\ref{tab:compliance}), and thus $B_c$ becomes less negative on heating with $\beta_c$ becoming more negative. Hence higher temperatures enhance NLC.

To understand this, we note that the values of $\alpha_{ab}$ and $\alpha_c$ depend on $\overline \gamma_{ab}$ and $\overline \gamma_{c}$ weighted by the compliances, as given in Eqs~\ref{linearcteab} and~\ref{linearctec}. Since the compliances listed in Table~\ref{tab:compliance} change little with pressure, any significant change of the CTE with pressure must be due to the change of the overall Gr\"{u}neisen parameters.

\begin{figure}[t]
\begin{center}
\includegraphics[width=0.48\textwidth]{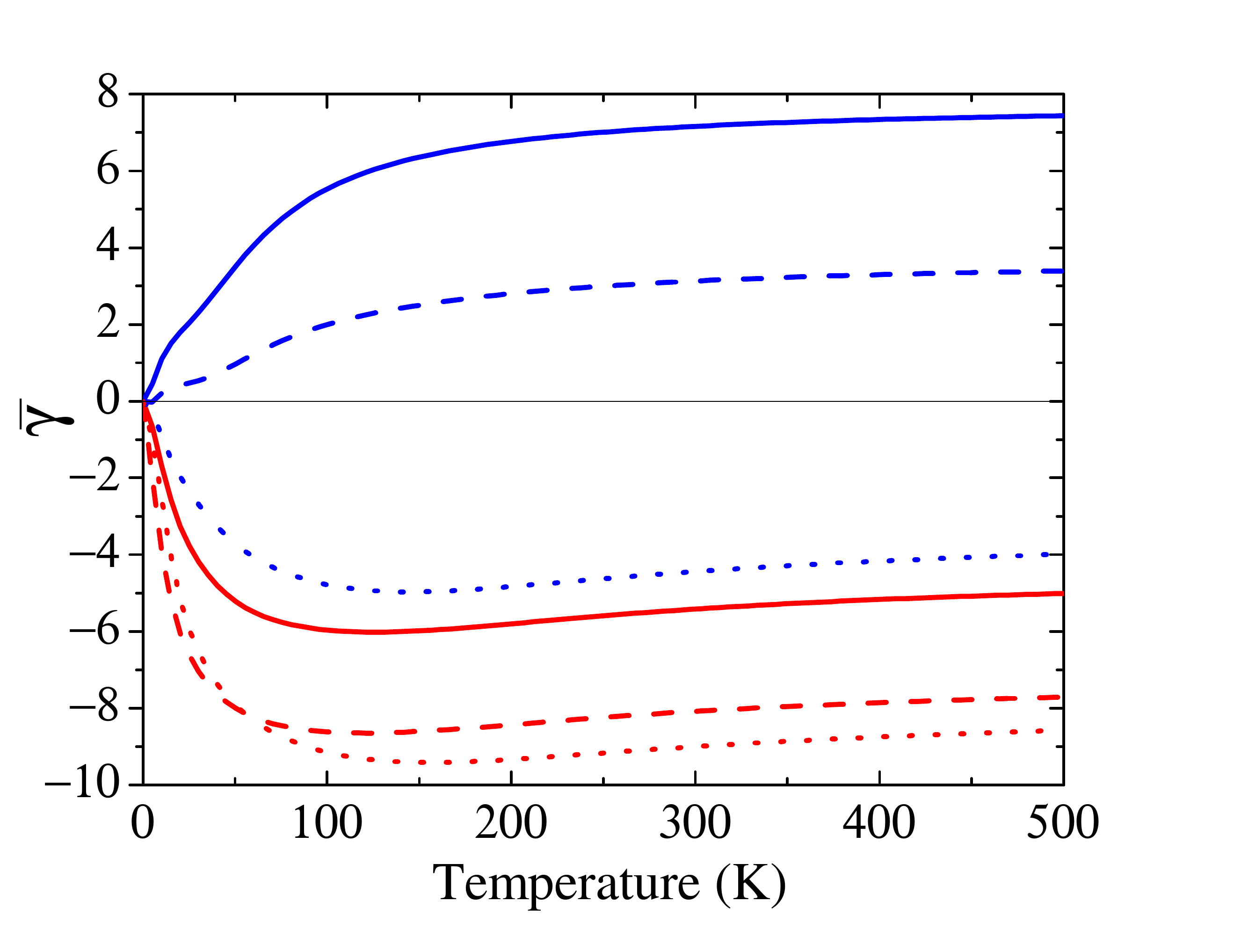}
\end{center}
\caption{\label{fig:overallgamma} Calculated overall Gr\"{u}neisen parameters $\overline \gamma_{ab}$ (in blue) and $\overline \gamma_{c}$ (in red) by Eq.~\ref{overallgamma}. $\overline \gamma_{ab}$ at 0.0, 0.04 and 0.1~GPa correspond to solid, dashed and dotted lines, respectively. $\overline \gamma_{c}$ at the same pressures corresponds to solid, dashed and dotted lines, respectively. From 0.0 to 0.1~GPa, $\overline \gamma_{ab}$ decreases significantly and even becomes negative at 0.1~GPa, while $\overline \gamma_c$ decreases much less.}
\end{figure}

In the temperature range of 0--500 K, only contributions from the low-frequency modes ($\leq 9$ THz) (Fig.~\ref{fig:dosgp}(a)) are important. At zero pressure, contributions from the low-frequency modes result in positive $\overline \gamma_{ab}$ and negative $\overline \gamma_c$ as shown in Fig.~\ref{fig:overallgamma}. Since $s_{11}$ and $s_{12}$ are positive and $s_{13}$ is negative, both $\overline \gamma_{ab}$ and $\overline \gamma_{c}$ would contribute constructively to the positive value of $\alpha_{ab}$ according to Eq.~\ref{linearcteab}. Similarly, since $s_{13}$ is negative and $s_{33}$ is positive, $\overline \gamma_{ab}$ and $\overline \gamma_{c}$ would also contribute constructively to the negative value of $\alpha_{c}$. Thus, the increase of $\overline \gamma_{ab}$ (becoming more positive) and decrease of $\overline \gamma_{c}$ (becoming more negative) would enhance the linear PTE and NTE, while the decrease of $\overline \gamma_{ab}$ and increase of $\overline \gamma_{c}$ would reduce the linear PTE and NTE of the material.

Fig.~\ref{fig:overallgamma} shows that there is a significant decrease of $\overline \gamma_{ab}$ and a smaller decrease of $\overline \gamma_{c}$ on compression. According to Eqs~\ref{linearcteab} and~\ref{linearctec}, the first effect is more dominant and results in the large decrease in the magnitude of both $\alpha_{ab}$ and $\alpha_c$ with pressure, corresponding to the conventional decrease of elastic moduli on heating ($\partial B_{ab}/\partial T \propto \partial \alpha_{ab}/\partial p < 0$) and the heat enhancement of NLC ($\partial B_{c}/\partial T \propto \partial \alpha_{c}/\partial p > 0$), respectively.

It is interesting to note this enhancement of NLC on heating could not happen without the hinging mechanism in the structure working efficiently, because it is this mechanism that gives almost the same magnitudes to $s_{13}$ and $s_{33}$ (as discussed in Section~\ref{groundstate}), which in turn provide the same weighting of $\overline \gamma_{ab}$ and $\overline \gamma_{c}$ in their contributions to $\alpha_c$. If we had the case where the hinging is not effective, a much smaller value of $s_{13}$ compared to $s_{33}$ would make the decrease of $\overline \gamma_c$ dominate, resulting in a decrease of $\alpha_c$ on compression (corresponding to $\partial B_{c}/\partial T \propto \partial \alpha_{c}/\partial p < 0$); in this case the enhancement of NLC on heating would not be observed.

\subsection{Exceptionally large $\partial \alpha_V/\partial p$}\label{softening}

Another interesting finding in Fig.~\ref{fig:cote} is the exceptionally large reduction in $\alpha_V$ on compression. The magnitude of $\partial \alpha_V / \partial p$ is found to be about 1125~MK$^{-1}$/GPa from 0.0 to 0.04~GPa and 2083~MK$^{-1}$/GPa from 0.04 to 0.1~GPa, values that are more than 10 times larger than what is normally considered as a large value~\cite{Cetinkol2008} (\textit{ca} 100~MK$^{-1}$/GPa) and more than 10$^4$ times larger than that of a hard metal~\cite{Fangmetal2010}.

From 0.0 to 0.1~GPa, the linear CTE of the material is reduced from its colossal value to a more moderate value of about $\pm25$~MK$^{-1}$ which is similar to the values found in the NTE metal cyanides~\cite{Goodwin2005,Fangmd2013}. As discussed in the previous section, such significant reduction in the magnitudes of $\alpha_{ab}$ and $\alpha_c$ is due to the large decrease of $\overline \gamma_{ab}$. In particular, when $\overline \gamma_{ab}$ becomes negative at 0.1 GPa, it begins to contribute to $\alpha_{ab}$ and $\alpha_c$ (Eqs~\ref{linearcteab} and~\ref{linearctec}) with opposite sign to that of $\overline \gamma_{c}$.

The significant decrease on compression of $\overline \gamma_{ab}$ is attributed to the large decrease in $\gamma_{ab}$ of most low-frequency modes ($\leq9$~THz), especially the modes with frequencies $<1.0$~THz (as those at wave vector A in Fig.~\ref{fig:phonon}). On one hand, such a decrease is related to the increase of mode frequencies (see Eq.~\ref{gammaa}) upon hydrostatic compression as shown in Fig.~\ref{fig:dosgp}(d). On the other hand, the sign change of $\gamma_{ab}$ at 0.1 GPa is indicated in Figs~\ref{fig:dosgp}(a) to (c) by the coloured DoS according to the values of $\gamma_{ab}$ at different pressures.

The sign change of $\gamma_{ab}$ of the low-frequency modes under pressure can be explained with the help of Fig.~\ref{fig:Amode1}. As discussed previously, the transverse vibration of the CN--Ag--NC bridge of such modes can pull the connected Co closer hence contract the $c$ dimension of the crystal. With relaxed Co--CN--Ag--NC--Co linkage at zero pressure, reducing the $a$ and $b$ dimensions of the unit cell tends to extend the $c$ dimension due to the hinging mechanism. This would make the transverse vibration that contracts the dimension more difficult and result in positive $\gamma_{ab}$ in Eq.~\ref{gammaa}. However, at high hydrostatic pressures, large elongation in the $c$ dimension (due to the giant NLC of the material) would largely extend the Co--CN--Ag--NC--Co linkage. This time, reducing the $a$ and $b$ dimensions with fixed $c$ of the unit cell can accommodate part of the extension in the linkage and make the linkage less taut. This would in turn make it easier for the CN--Ag--NC linkage to vibrate transversely, which would result in negative $\gamma_{ab}$ in Eq.~\ref{gammaa}.

The scissor-like behaviour of the change of linear CTE seen in the upper panel of Fig.~\ref{fig:cote}, namely the decrease of $\alpha_{ab}$ accompanied by the increase of $\alpha_c$ upon compression, makes the combined $\alpha_V$ in Eq.~\ref{linearctev} close to zero at high pressure. The large value of $s_{11}$ due to the weak interaction between Ag atoms in the $a$--$b$ plane makes sure that the contribution from $\overline \gamma_{ab}$ to $\alpha_{ab}$ in Eq.~\ref{linearcteab} dominates, so that $\alpha_{ab}$ would decrease largely according to the decrease of $\overline \gamma_{ab}$. On the other hand, as discussed in the previous section, the effective hinging mechanism guarantees the similarly large increase of $\alpha_c$. Thus, it is the dispersive interaction together with the hinging mechanism that make $\alpha_{ab}$ and $\alpha_c$ change with pressure like a scissor.

\begin{figure}[t]
\begin{center}
\includegraphics[width=0.48\textwidth]{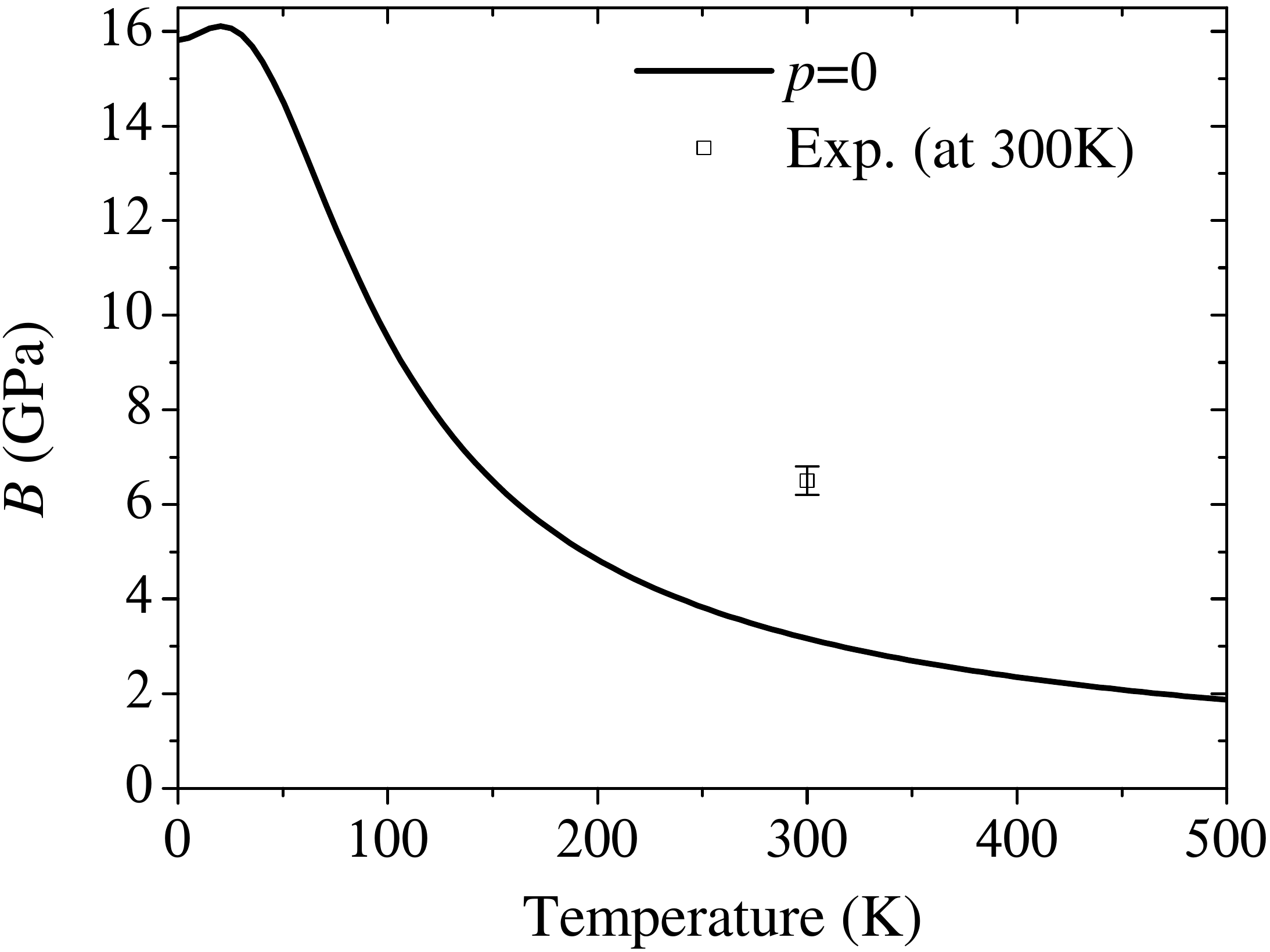}
\end{center}
\caption{\label{fig:bt} Calculated temperature dependence of the bulk modulus $B$ of Ag$_3$Co(CN)$_6$ at zero pressures using Eq.~\eqref{bulkmodulustemperature}. The great softening of $B$ on heating brings the calculated value in better agreement to the experimental one at room temperature.}
\end{figure}

According to the relation~\cite{Fangexp2013}
\begin{eqnarray}\label{bsoftening}
\left( {\frac{{\partial B_V }}{{\partial T}}} \right)_p  = B_V^2 \left( {\frac{{\partial \alpha _V }}{{\partial p}}} \right)_T,
\end{eqnarray}
the giant reduction of $\alpha_V$ with pressure implies a giant decrease of $B$ on heating. From Eq.~\ref{bsoftening}, $B(T)$ can be calculated as
\begin{eqnarray}\label{bulkmodulustemperature}
B(T) = \left( {\frac{1}{{B_{T = 0} }} - \int_0^T {\frac{{\partial \alpha }}{{\partial p}}} dT} \right)^{ - 1},
\end{eqnarray}
and is shown in Fig.~\ref{fig:bt}. From 0.0 to 300 K, $B$ is reduced by $75\%$ which is much larger than the observed giant softening ($\sim 45\%$) of the isotropic NTE material ZrW$_2$O$_8$~\cite{Pantea2006} on heating. Such softening results in a value of $B$ in much better agreement with the experimental value of 6.5(3) GPa at room temperature~\cite{GoodwinAgCoCN2008}, as shown in Fig.~\ref{fig:bt}.

\section{Conclusions}

By including the dispersive correction in the DFT GGA calculation, we are now able to correctly reproduce the ground state of Ag$_3$Co(CN)$_6$ as well as the the high-pressure phase of the material having the interdigitated structure.

We found that, by using the DFPT+D calculated phonons, it is almost the same set of low-frequency modes that contribute to both linear PTE and NTE of the material with their linear Gr\"{u}neisen parameters showing similar magnitudes but with opposite sign. Such modes, as those around the wave vector A and the middle point along the H$\rightarrow$K, correspond to the transverse vibrations of the CN--Ag--NC bridge within the Co--CN--Ag--NC--Co linkage that can transfer the expansion in the $a$($b$) dimension to the contraction in the $c$ dimension.

From the DFPT+D results, we have predicted that the value of $\alpha_c$ of Ag$_3$Co(CN)$_6$ increases on compression, contrary to what is normally seen in PTE and NTE materials. In turn this suggests that the NLC of Ag$_3$Co(CN)$_6$ will be enhanced on heating. We also predicted an exceptionally large reduction in volume CTE on compression, which corresponds to the change of sign of the linear Gr\"{u}neisen parameters under pressure together with the right elasticity of the material. The latter is based on the weak interactions between Ag atoms in the $a$--$b$ plane and the effective hinging mechanism in the structure. This property also suggests a giant softening of the material on heating with a reduction in the bulk modulus of about $75\%$ from 0--300 K.

The method and results presented in this work would be able to apply to other framework materials, such as KMn[Ag(CN)$_2$]$_3$ and Zn[Au(CN)$_2$]$_2$, that have atoms (e.g. Ag and Au) with large dispersive interactions and show large anisotropic properties of PTE/NTE as well as NLC~\cite{Cairns2012,Kamali2013,Cairns2013,Gatt2013}. It would be interesting in a future study to see if the phenomena of heat enhancement of NLC and giant reduction of volume CTE on compression predicted for Ag$_3$Co(CN)$_6$ can also be found in these other materials. It would be also interesting to use other schemes to include the van der Waals dispersion correction (such as the use of non-local Langreth-Lundqvist functional~\cite{Dion2004} in the DFT) in calculating properties of these materials and compare the results.

\begin{acknowledgements}
We gratefully acknowledge financial support from the Cambridge International Scholarship Scheme (CISS) of the Cambridge Overseas Trust and Fitzwilliam College of Cambridge University (HF). We thank the CamGrid high-throughput environment of the University of Cambridge. We thank the UK HPC Materials Chemistry Consortium, funding by EPSRC (EP/F067496), to allow us to use the HECToR/ARCHER national high-performance computing service provided by UoE HPCx Ltd at the University of Edinburgh, Cray Inc and NAG Ltd, and funded by the Office of Science and Technology through EPSRC's High End Computing programme.
\end{acknowledgements}

\end{document}